\documentclass[aps,twocolumn,showpacs,superscriptaddress]{revtex4-1}
\usepackage{amsfonts}
\usepackage{amsmath}
\usepackage{amssymb}
\usepackage{lipsum}
\usepackage{graphicx}
\usepackage{bm}
\usepackage{natbib}
\usepackage{float}
\usepackage{multirow}
\usepackage{ulem}
\usepackage{hyperref}
\usepackage{comment}
\hypersetup{colorlinks=true,
            linkcolor=blue,
            citecolor=blue,
            urlcolor=orange}
        
\usepackage[dvipsnames]{xcolor}

\usepackage{comment}

\begin{document}
\preprint{APS/123-QED}
\title{Effect of vacancies on magnetic correlations and conductance in graphene nanoflakes with realistic Coulomb interaction}
\author{V. S. Protsenko}
\affiliation{M. N. Mikheev Institute of Metal Physics of the Ural Branch of the Russian Academy of Sciences, S. Kovalevskaya Street 18, 620990 Yekaterinburg, Russia}
\author{A. A. Katanin}
\affiliation{Center for Photonics and 2D Materials, Moscow Institute of Physics and Technology, Institutsky lane 9, Dolgoprudny, 141700, Moscow region, Russia}
\affiliation{M. N. Mikheev Institute of Metal Physics of the Ural Branch of the Russian Academy of Sciences, S. Kovalevskaya Street 18, 620990 Yekaterinburg, Russia}
\date{\today}
\begin{abstract}
We study the effect of various configurations of vacancies on the magnetic properties of graphene nanoflake (GNF) with screened realistic long-range electron interaction [T. O. Wehling, {\it et al.}, Phys. Rev. Lett. {\bf 106}, 236805 (2011)] within the functional renormalization group approach. 
In agreement with previous studies, the presence of vacancies in GNF yields a strong enhancement of spin-density-wave (SDW) correlations. We show however that only some part of the considered configurations of vacancies posses SDW ground state. 
The probability of a system with a random configuration of vacancies to be in the SDW ground state increases with increase of vacancy concentration. The disorder-averaged sublattice magnetization increases linearly with the concentration of vacancies.
The ratio of the sublattice magnetizations at the center and edges of GNF, averaged over various realizations of disorder, depends only weakly on the number of vacancies. The effects of vacancies on the linear conductance and charge properties of GNF are discussed.
\end{abstract}
 \maketitle
\section{Introduction} 

Graphene nanoflakes (GNFs) are nanostructures formed by a small piece of graphene sheet~\cite{Snook_2011}.
As in graphene, both short-range (local) and long-range electron-electron interactions play a crucial role in magnetic, charge, and electronic transport properties of graphene nanoflakes~\cite{DCA, Valli_2016, Valli_2018, Yazyev_2010, Fernandez_2007, Kabir_2014, Yamashiro_2003, Zhu_2006, Chacko_2014, Luo_2014, Wunsch_2008, Sheng_2013, GNF_2021, Valli_2019}. In particular, the interplay between local and long-range electron correlations leads to the appearance of spin-density wave (SDW) and charge-density wave (CDW) correlations~\cite{DCA, Chacko_2014, Valli_2019, GNF_2021, Yamashiro_2003}. 
For realistic parameters of Coulomb interaction, the pristine graphene and GNF without disorder appear to be in the semimetal (SM) phase, which does not possess charge or magnetic order~\cite{Ulybyshev_2013, GNF_2021}. This also applies to GNFs of various sizes and boundary geometries~\cite{GNF_2021}. 

At the same time, it has been established from both experiments and theory that the disorder can lead to the appearance of magnetism in graphene and graphene-based nanostructures. It has been reported that for disorder created by 
vacancies, local magnetic moments~\cite{Vozmediano_2005, Pereira_2008, Pereira_2006, Ugeda_2010, Chen_2011} and a pronounced magnetic order~\cite{Yazyev_2007, Palacios_2008, Droth_2015} can appear. This vacancy-induced magnetism has been studied in graphene and various graphene nanostructures, including GNFs. In particular, the phase transition to an antiferromagnetic state, driven by the presence of vacancies or adatoms, has been investigated for large graphene clusters with the realistic long-range Coulomb interaction~\cite{{Ulybyshev_2015}}. It is shown that the effects of the realistic long-range interaction result in the appearance of magnetic moments near adatoms, leading to an antiferromagnetic ordering of graphene clusters at low temperatures.

{In the previous study of vacancies in large graphene clusters with realistic Coulomb interaction, for each particular cluster, one configuration with a random uniform distribution of a fixed concentration of vacancies was considered~\cite{Ulybyshev_2015}. At the same time, finite size of GNFs leads to the appearance of edge states which have properties different from those of bulk states. For pristine GNFs, sites located near the edges (or at the edges) develop a more pronounced magnetic order or contribute to the formation of magnetic moments. Consequently, one can expect that vacancies located at or close to the edges of the system have a different impact on the state of the system than vacancies located far from the edges. Due to this, the average over various disorder configurations in GNF systems 
may be especially important. Also, analyzing various disorder configurations may yield information on the possible range of magnetic properties, which can be produced by the presence of defects or vacancies in GNF.
} 

To analyze the properties of
GNF, the functional renormalization group (fRG) approach~{\cite{Salmhofer_1, Metzner}} which was previously used to describe pristine GNF systems~{\cite{GNF_2021}}, was shown to be a useful tool. The application of this method for exploring the phase diagram of pristine GNF systems~\cite{GNF_2021} demonstrates agreement with sophisticated many-body numerical approaches, such as the dynamical cluster approximation (DCA), the dynamic mean-field theory (DMFT), and the hybrid quantum Monte Carlo (QMC) simulations. This approach can be applied to study the effects of disorder, since it allows relatively fast study of many different disorder configurations. 

In this paper, we study how the disorder caused by the presence of vacancies manifests itself in the magnetic, charge, and transport properties of a GNF at zero temperature. We consider the realistic parameters of electron-electron interactions in graphene~\cite{Wehling_2011}, which account for the screening of the interaction by $\sigma$ orbitals and have been determined by accurate first-principles calculations. The difference between the realistic screened Coulomb potential and the standard $1/r$ Coulomb potential leads to a significant shift in the critical value of interaction for the SM-CDW phase transitions for graphene~\cite{Ulybyshev_2013,GNF_2021}. 

Considering various realizations of disorder {and extrapolating the results to zero staggered magnetic field}, we show that, for the same number of vacancies, the GNF can be either in the SM phase or in the SDW phase, depending on the spatial distribution of vacancies.
By averaging over ensembles of independent configurations, we show that the presence of vacancies leads to a strong enhancement of SDW magnetic correlations. Associated with this enhancement, linear increase of the disorder-averaged relative staggered magnetization {of the whole system} (as well as staggered magnetization at the center and edges of the GNF) with the vacancy concentration is accompanied by the weakly nonlinear behavior of the conductance as a function of vacancy concentration.
 At the same time, we find that vacancies do not lead to the charge order of the system.


\section{Model and method}

\begin{figure}[t]
\centering
\includegraphics[width=0.77\linewidth]{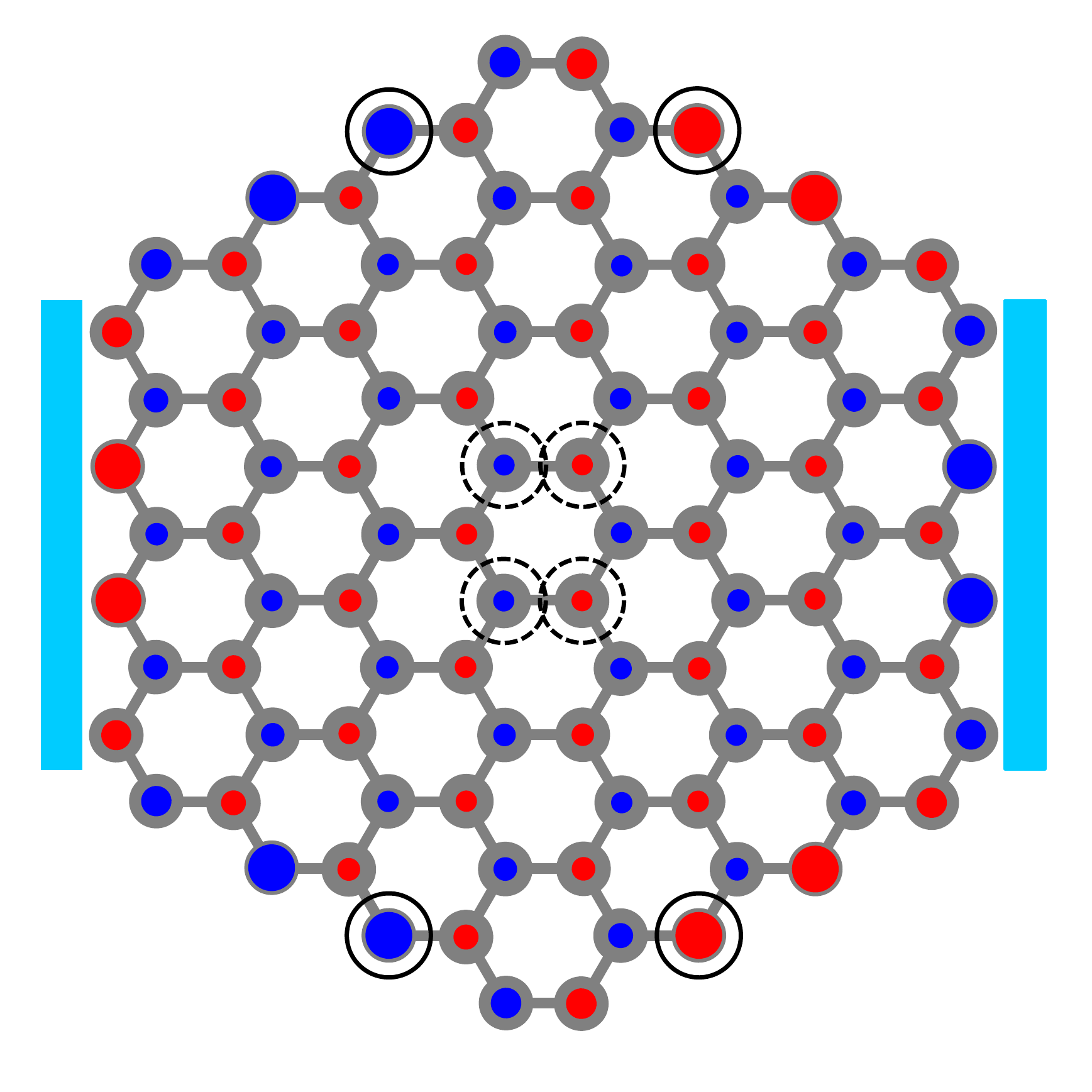}
\caption{(Color online) Zigzag-edge GNF system with $96$ atoms (gray dots) connected to two leads (rectangles). The red (blue) dots correspond to $A$ ($B$) sublattices with positive (negative) $m_{j}=\langle n_{j,\uparrow}-n_{j,\downarrow}\rangle$; their size is proportional to the on-site magnetization $|m_j|$, which is induced by the weak staggered magnetic field $h=0.0185t$. The open (dashed open) circles indicate the sites corresponding to the $\max\{|m_{j}|\}\approx 0.043$ ($\min\{|m_{j}|\}\approx 0.019$). The relative staggered magnetization $S^{(0)}_{\rm st}\approx 0.024$.}
\label{ConfVac0}
\end{figure}

The system under consideration is illustrated in Fig.~\ref{ConfVac0}. It consists of a zigzag-edge GNF with $N_{\rm at}=96$ atoms connected to two equivalent metallic leads. The total Hamiltonian is $\mathcal{H}= \mathcal{H}_{\rm GNF}+\mathcal{H}_{\rm leads}+\mathcal{H}_{\rm T}$. The first term describes the isolated GNF,
\begin{multline}
 \mathcal{H}_{\rm GNF}=\sum_{\sigma}\sum_{i\in A}\epsilon_{\sigma}^{A}n_{i,\sigma}+\sum_{\sigma}\sum_{i\in B}\epsilon_{\sigma}^{B}n_{i,\sigma}\\-t\sum_{<ij>,\sigma}d^{\dagger}_{i,\sigma}d_{j,\sigma}+\dfrac{1}{2}\sum_{i,j}U_{ij}\left(n_{i}-1\right)\left(n_{j}-1\right).
 \label{GNFHamiltonian}
\end{multline}
Here, $d^{\dagger}_{i,\sigma}$ $\left(d_{i,\sigma}\right)$ is a creation (annihilation) operator of an electron at the lattice site $i$ {of $A$ or $B$ sublattice} with a spin index $\sigma=\pm 1/2$ (or $\sigma=\uparrow,\downarrow$), $n_{j,\sigma}=d^{\dagger}_{j,\sigma}d_{j,\sigma}$, and $n_{j}=n_{j,\uparrow}+n_{j,\downarrow}$. The on-site energy parameters are chosen to be $\epsilon_{\sigma}^{A(B)}=\pm\left(\delta-h\sigma\right)$, the parameter $\delta$ and {staggered magnetic field} $h$ are introduced in order to explicitly break the spin and sublattice symmetry of the GNF, $t=2.7$~eV 
is the nearest-neighbor hopping parameter, and summation in the third term of Eq.~(\ref{GNFHamiltonian}) is taken over nearest neighbor sites. The last term in Eq.~(\ref{GNFHamiltonian}) describes the electron-electron interactions with the potential $U_{ij}$ that includes both on-site $U=U_{ii}$ and nonlocal $U_{i\ne j}$ contributions. For the parameters $U_{ij}$ up to the third-nearest-neighbors $r_{ij}\leq r_{03}=2a$ ($a=0.142$ nm is graphene's lattice constant) we use the realistic values for graphene \cite{Wehling_2011}, which were determined within constrained random phase approximation (cRPA)  (see Table~1 of Ref.~\cite{Wehling_2011} for the corresponding values of $U_{ij}$).  
This potential accounts for the realistic screening of Coulomb potential by electrons on $\sigma$ orbitals, which, in particular, is essential for a consistent description of the SM-SDW phase transition in monolayer graphene~\cite{Ulybyshev_2013}. At distances larger than the distance between third-nearest-neighbor lattice sites $r_{ij}>r_{03}$ the realistic potential is approximated by $U_{ij}=1/(\epsilon_{\rm eff} r_{ij})$ with effective dielectric permittivity {$\epsilon_{\rm eff}=1/(U_{03}r_{03})\approx 1.41$}, as in Ref.~\cite{Ulybyshev_2013}. 

 The term $\mathcal{H}_{\rm leads}$ describes 1D semi-infinite Fermi-liquid leads which are assumed to be equivalent and are modeled by
\begin{eqnarray}
\mathcal{H}_{\rm leads}&=&-\sum_{\alpha=L,R}\sum_{k=0}^{\infty}\sum_{\sigma}\left[\mu_{\alpha}c^{\dagger}_{\alpha,k,\sigma}c_{\alpha,k,\sigma}\right.\notag\\
&+&\left.\tau(c^{\dagger}_{\alpha,k+1,\sigma}c_{\alpha,k,\sigma}+\text{H.c.})\right],
\label{H_lead}
\end{eqnarray}
where $c^{\dagger}_{\alpha,k,\sigma}(c_{\alpha,k,\sigma})$ is a creation (annihilation) operator for an electron on the $k$ lattice site of the left $\alpha=L$ or right $\alpha=R$ lead, $\tau$ denotes  nearest-neighbor hopping between the sites of the leads and $\mu_{\alpha}$ is the chemical potential.
Finally, $\mathcal{H}_{\rm T}=-\sum_{\sigma,\alpha,i_\alpha}(V^{\phantom{\dagger}}_{i_\alpha, \alpha} c^{\dagger}_{\alpha,0,\sigma}d^{\phantom{\dagger}}_{i_\alpha,\sigma}+\text{H.c.})$ describes the hopping between GNF and the leads, where $V_{i,\alpha}$ is the coupling matrix element between the $i$th site of the GFN and the last site of the lead $\alpha$, and summation is performed over sites $i_\alpha$ that are closest to the lead $\alpha$. Note that the parameters of the leads and the coupling matrix elements $V_{i,\alpha}$ only indirectly influence the GNF's degrees of freedom through the hybridization function~\cite{Wingreen_1994, Enss_thesis} $\Gamma_{ij}^{\alpha}\left(\epsilon\right) = 2\pi {V_{i,\alpha}V^{*}_{j,\alpha}\rho_{\rm leads}}$, where $\rho_{\rm leads}$ is the local density of the states of the leads. We assume equal GNF-leads couplings $V_{i,,\alpha}=V$ and consider the wide-band limit approximation~\cite{Verzijl_2013, Wingreen_1994}. This leads to an energy independent hybridization strength $\Gamma$~\cite{GNF_2021}, which absorbs all the information about the noninteracting environment.

To model the effect of disorder caused by vacancies, we randomly remove a fixed number of atoms $N_{\rm vac}$ from the system shown in Fig.~\ref{ConfVac0}. Sites that are directly connected to the leads are not removed. In order to maintain the half-filling of the system,
which in the presence of $N_{\rm vac}$ vacancies corresponds to $N=N_{\rm at}-N_{\rm vac}$ electrons in the system, an equal number of atoms is removed from each sublattice.
To detect the formation of SDW and CDW phases we calculate the relative staggered magnetization
\begin{equation}
S^{(N_{\rm vac})}_{\rm st} =\left(\langle N_{A,\uparrow} \rangle+\langle N_{B,\downarrow}\rangle-\langle N_{A,\downarrow} \rangle -\langle N_{B,\uparrow} \rangle\right)/N
\label{S_eq}
\end{equation}
and the relative difference between the occupation of $A$ and $B$ sublattices
\begin{equation}
\Delta^{(N_{\rm vac})}_{\rm st}=\left(\langle N_{B,\uparrow} \rangle+\langle N_{B,\downarrow}\rangle-\langle N_{A,\uparrow} \rangle -\langle N_{A,\downarrow} \rangle\right)/N,
\label{Delta_eq}
\end{equation}
which can be considered as SDW and CDW order parameters, where ${\langle}N_{A(B),\sigma}{\rangle}=\sum_{i\in A(B)} {\langle}n_{i,\sigma}{\rangle}$,  $\langle n_{j,\sigma}\rangle$ 
is the average occupation of a lattice site $j$ for spin $\sigma$. Within the above definitions, the maximal SDW (CDW) order parameter is 
$S^{(N_{\rm vac})}_{\rm st}=1$ ($\Delta^{(N_{\rm vac})}_{\rm st} =1$). In contrast, in the SM phase, Eqs.~(\ref{S_eq})-(\ref{Delta_eq}) yield $S^{(N_{\rm vac})}_{\rm st}=0$ and $\Delta^{(N_{\rm vac})}_{\rm st} =0$, indicating the absence of magnetic and charge order.

We consider below  the temperature $T=0$. 
We also study the linear conductance, which is given at $T=0$ by Landauer formula~\cite{Valli_2019, Datta_1995, Oguri_2001}
\begin{equation}
G^{(N_{\rm vac})}=4\Gamma^{2}{G_{0}} \sum_{i,j,\sigma}{\left|\mathcal{G}^{}_{ij,\sigma}({\omega \to 0^{+}})\right|^{2}}. 
\label{G_eq}
\end{equation}
where $\mathcal{G}^{}_{ij,\sigma}(\omega)$ is the Green's function of electron moving from site $i$ to site $j$ and summation over site indexes is restricted to sites of GNF which are connected to the left (right) lead (see Fig.~\ref{ConfVac0}). Equation~(\ref{G_eq}) is valid also in the presence of the interaction due to vanishing vertex corrections in the $T=0$ limit~\cite{Oguri_2001, Enss_thesis}, $G_{0}=e^{2}/h$ is the conductance quantum per spin projection. Note that Eq.~(\ref{G_eq}) assumes only local hybridization processes~\cite{Valli_2019}. At $T=0$ we also have $\langle n_{j,\sigma}\rangle=\int{d\omega e^{i\omega 0^{+}}
	\mathcal{G}
	_{jj,\sigma}}(i \omega)/{2\pi}$.
 As follows from Eqs.~(\ref{S_eq})--(\ref{G_eq}), $S^{(N_{\rm vac})}_{\rm st}$, $\Delta^{(N_{\rm vac})}_{\rm st}$, and $G^{(N_{\rm vac})}$ are determined by the Green's function $\mathcal{G}_{\sigma}(i \omega)$. We calculate the Green's function of the system by using the functional renormalization group (fRG) method~\cite{Salmhofer_1, Metzner} within the coupled ladder approximation~\cite{Bauer_2014,Weidinger_2017} {and the reservoir frequency cutoff scheme~\cite{Karrasch_2010}}. The implementation of this method follows the same fRG procedure as in Ref.~\cite{GNF_2021} [see Eqs.~(8)--(19) of that paper]. The leads contribution to the Green's function is given by $\Sigma_{\rm leads}=-i\Gamma$~\cite{GNF_2021}.

To obtain statistically converged results, we perform an averaging over $n$ random realizations of $N_{\rm vac}$ vacancies in the GNF system defined as 
$\langle X^{(N_{\rm vac})} \rangle_{n}=({1}/{n})\sum_{k=1}^{n}{X^{(N_{\rm vac})}_{k}}$,
where $X^{(N_{\rm vac})}_{k}$ is a value of $X^{(N_{\rm vac})}\in\{S^{(N_{\rm vac})}_{\rm st}, \Delta^{(N_{\rm vac})}_{\rm st}, G^{(N_{\rm vac})}\}$ obtained for $k$th random disorder realization. In the following calculations we set $\Gamma=0.02t$ and {$\delta=h$}.\par

\section{Magnetic phases in the presence of disorder}

 \begin{figure}[b]
\centering
\includegraphics[width=0.85\linewidth]{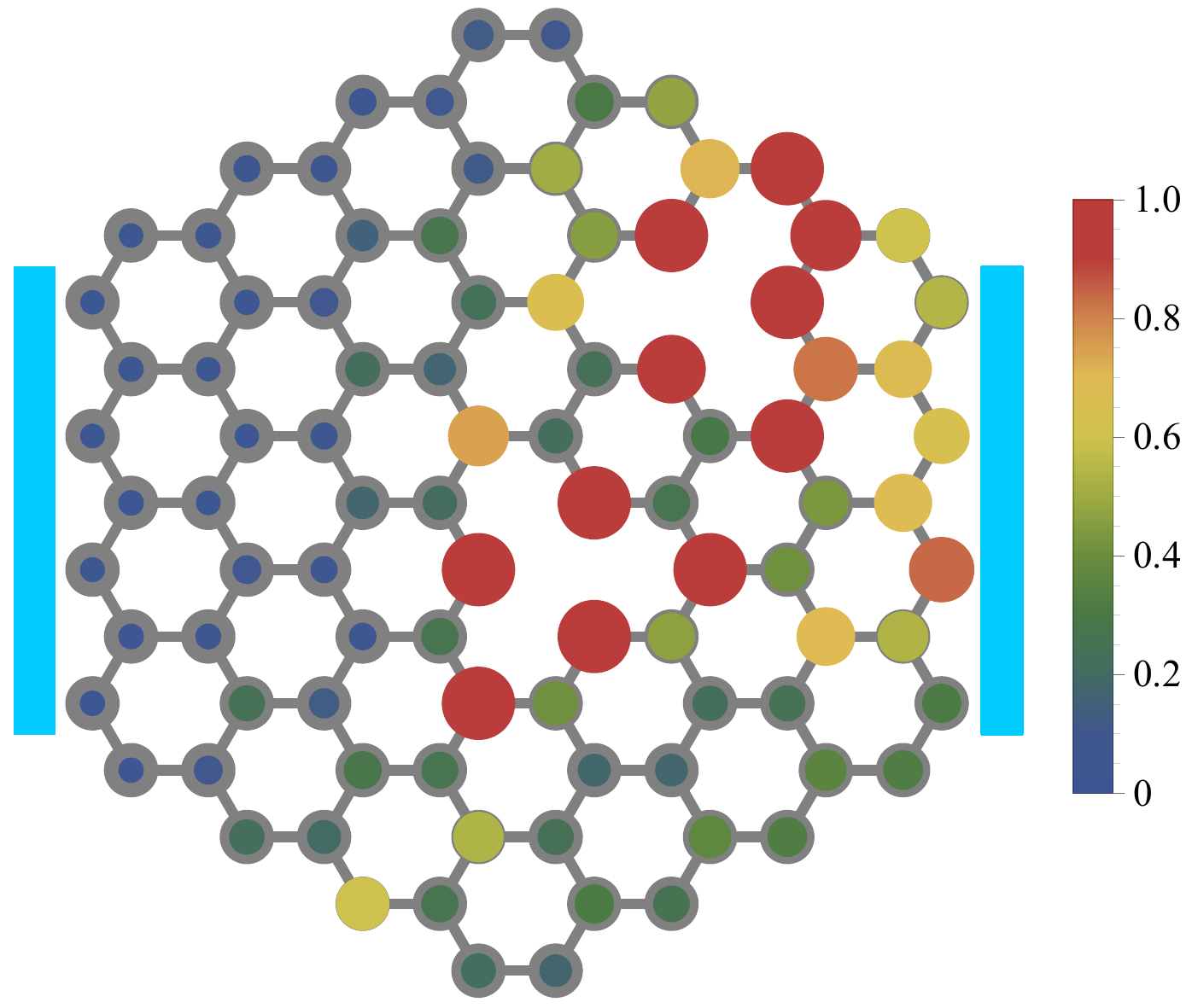}
\caption{(Color online) The distribution of the relative magnetization $\tilde{m}_{j}=m_j/m^{(0)}_j$ ($m^{(0)}_j$ is the magnetization of the $j$th site for $N_{\rm vac}=0$; see Fig.~\ref{ConfVac0}) in the system for $N_{\rm vac}=2$ for a given configuration of vacancies. The size of the colored dots is proportional to $\tilde{m}_{j}$ and the color corresponds to $\left(\tilde{m}_{j}-I_{\rm min}\right)/\left(I_{\rm max}-I_{\rm min}\right)$. Here $I_{\rm max}=\max\{\tilde{m}_{j}\}\approx 8.8$ and $I_{\rm min}=\min\{\tilde{m}_{j}\}\approx 1.2$. The relative staggered magnetization $S^{(2)}_{\rm st}\approx 2.4 S^{(0)}_{\rm st}$. The other system parameters 
are the same as in Fig.~\ref{ConfVac0}.}
\label{ConfVac2}
\end{figure}
{In this section, we analyze the response of the GNF system to a magnetic field and show that for a fixed number of vacancies $N_{\rm vac}$, depending on their configuration, the system belongs to either the SM or SDW ground state. We also perform subsequent average over various disorder configurations.
\subsection{Spatial distribution of the magnetization in the presence of vacancies}

\begin{figure}[b]
\centering
\includegraphics[width=0.9\linewidth]{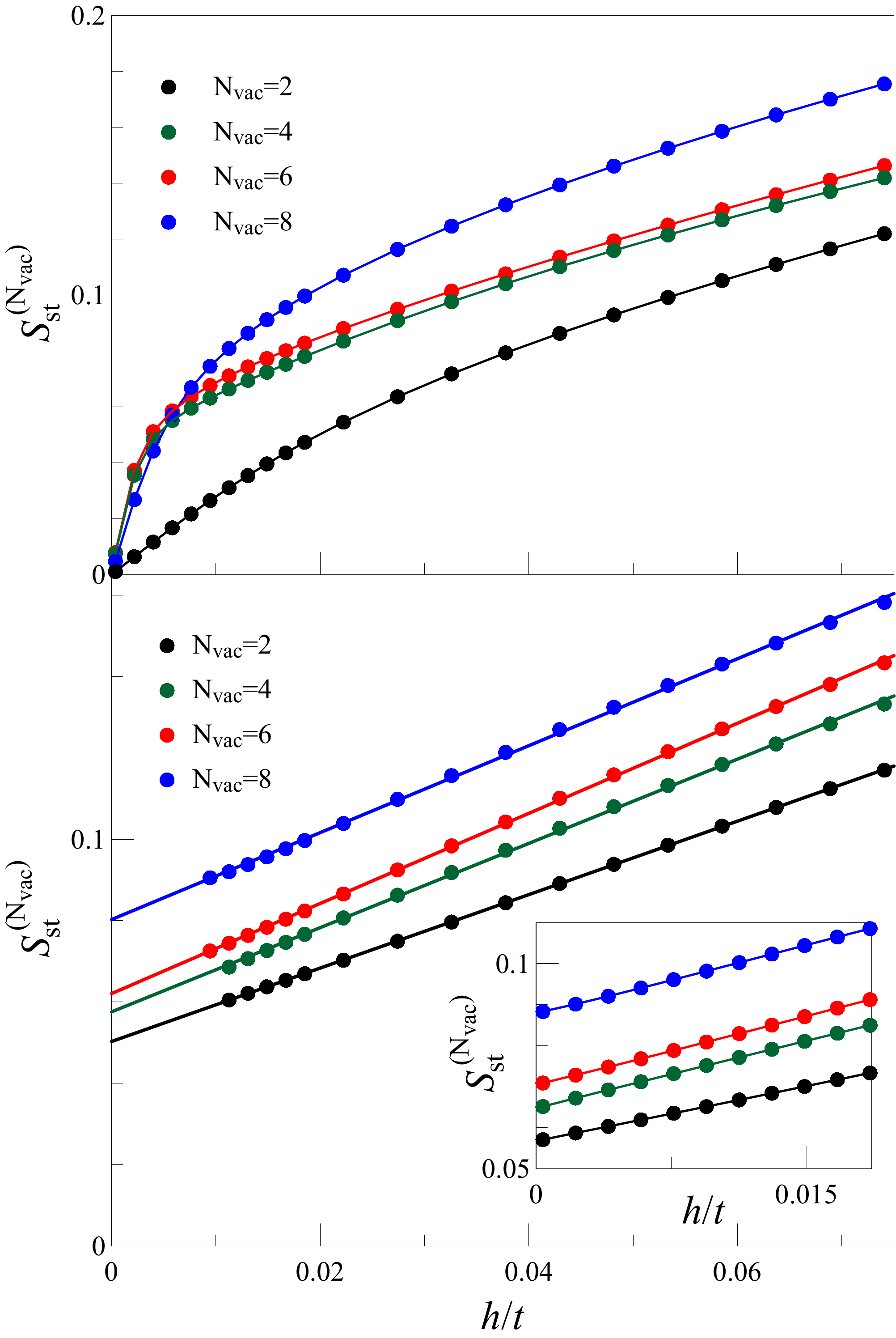}
\caption{(Color online) Examples of the magnetic-field $h$ dependence of the relative staggered magnetization $S_{\rm st}^{(N_{\rm vac})}$ for disorder realizations, for which $S_{\rm st}^{(N_{\rm vac})}$ vanishes (upper panel) or remains finite (lower panel) in the limit $h \rightarrow 0$, corresponding to the SM or SDW state. The inset in the lower panel displays the results of the fRG approach with the counterterm for the systems shown in the main part of the figure.}
\label{SzHp1}
\end{figure}

Let us first consider the results for the GNF system without vacancies ($N_{\rm vac}=0$). According to Ref.~\cite{Ulybyshev_2013}, suspended graphene with the realistic Coulomb potential is in the SM phase, {which corresponds to $S^{(0)}_{\rm st} \to 0$ for $h \to 0$}. 
Figure~\ref{ConfVac0} shows the distribution of the magnetization $m_j=\langle n_{j,\uparrow}-n_{j,\downarrow} \rangle$ in the absence of vacancies, which reflects the distribution of the magnetic response to a weak staggered magnetic field. One can see that the sites on the edges are characterized by larger response 
$|m_j|$. Note that the GNF-leads hybridization slightly suppresses the magnetization of the sites which are in a direct contact with the leads. In particular, for this reason, the sites that have the highest $|m_j|$ (marked by open circles) are located on the edges, which do not have a connection to the leads. In contrast, the central sites of the GNF show the lowest magnetic response.
\par
The distribution of the magnetization of one random configuration with $N_{\rm vac}=2$ is presented in Fig.~\ref{ConfVac2}. Compared to the case without vacancies (see Fig.~\ref{ConfVac0}), the magnetization of the sites around the vacancies is strongly enhanced. As a result, the relative staggered magnetization $S^{(2)}_{\rm st}$ for $N_{\rm vac}=2$ is substantially higher than in the case without vacancies $S^{(0)}_{\rm st}$, $S^{(2)}_{\rm st}\approx 2.4 S^{(0)}_{\rm st}$. We found qualitatively the same effects of vacancies on the distribution of the magnetization and relative staggered magnetization for systems with $N_{\rm vac} \leq 8$.\par
\subsection{The disorder configurations, possessing SDW and SM state}

\begin{figure}[b]
\centering
\includegraphics[width=1.0\linewidth]{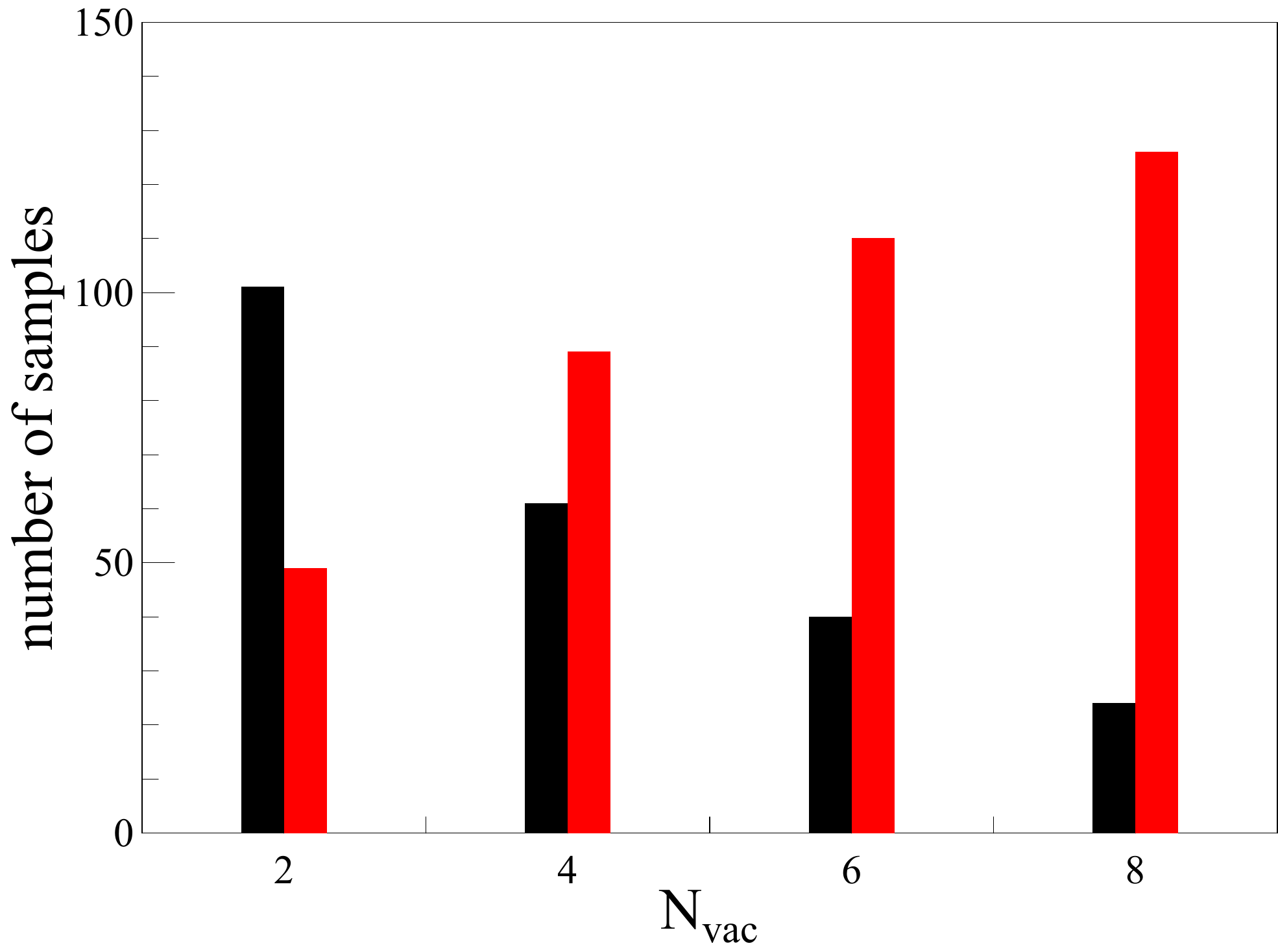}
\caption{(Color online) Bar graph representation of the distribution of SM (black) and SDW (red) phases for various even values of $N_{\rm vac}$.}
\label{SM_SDW_distr}
\end{figure}
{To determine the ground state of the GNF system in the presence of vacancies we analyze magnetic field dependences of the relative staggered magnetization $S_{\rm st}^{(N_{\rm vac})}\left(h\right)$ for different random configurations of $N_{\rm vac}$ ($N_{\rm vac}=2,4,6,8$) vacancies in the system. Considering the $S_{\rm st}^{(N_{\rm vac})}$ in the limit of zero magnetic field $h\to 0$, in this case, allows one to strictly separate the SM ($S_{\rm st}^{(N_{\rm vac})}\rightarrow 0$) and SDW ($S_{\rm st}^{(N_{\rm vac})}\neq 0$) phases.} 

\begin{figure*}[t]
\centering
\includegraphics[width=1.0\linewidth]{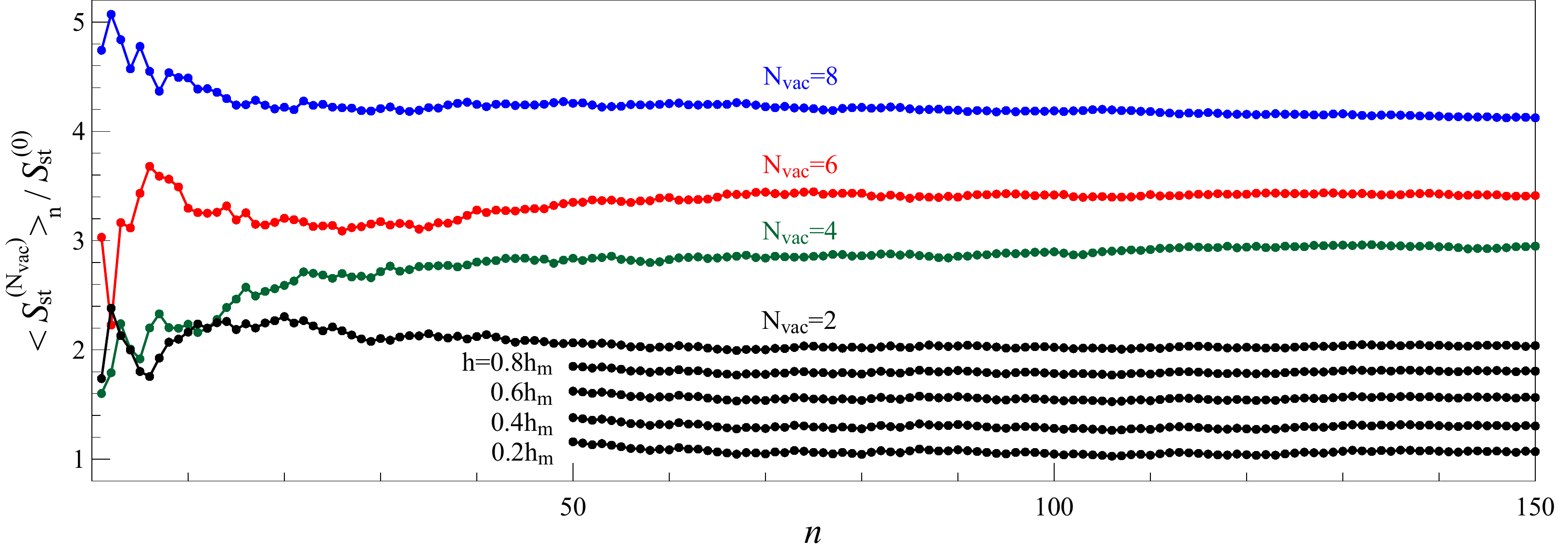}
\caption{(Color online) Upper four curves: The disorder-averaged relative staggered magnetization $\langle S_{\rm st}^{(N_{\rm vac})}\rangle_{n}$ with respect to {$S^{(0)}_{\rm st}$($h=h_{m}$)} as a function of the number of independent samples $n$ ($n\leq 150$) for $h=h_{m}$ and various $N_{\rm vac}$ (see labels at the dependencies). Lowest four dependencies show the data for $N_{\rm vac}=2$, $n{\geq}50$ and the magnetic fields  $h=0.8h_{m}$, $0.6h_{m}$, $0.4h_{m}$, and $0.2h_{m}$ (from the upper to the lower curve).}
\label{S_vac_n_p1}
\end{figure*}

{For each considered $N_{\rm vac}$, we find that there are two distinct behaviors of $S_{\rm st}^{(N_{\rm vac})}\left(h\right)$ that are possible depending on the position of vacancies in the system. The first type of behavior is characterized by $S_{\rm st}^{(N_{\rm vac})}\rightarrow 0$  
and relative staggered magnetization decreasing nonlinearly with decreasing $h$ 
(see examples 
in the upper panel of Fig.~\ref{SzHp1}), which 
corresponds to the SM state of the system. Apart from this, 
there are configurations which have linear dependencies of the relative staggered magnetization on $h$ and for which the self-energy diverges at small $h$ during the fRG flow and the sublattice magnetization, extrapolated to the limit $h\rightarrow 0$ does not vanish (see examples in the lower panel of Fig.~\ref{SzHp1}). This is an indication of the SDW state, which is inaccessible for our fRG approach for small magnetic fields.
To verify that these configurations correspond to SDW order, which is characterized by a nonzero staggered magnetization, we have also applied the counterterm extension of the fRG approach, corresponding to auxiliary magnetic field $\tilde{h}=0.03t$, switched off linearly with cut-off parameter $\Lambda$ starting from the scale $\Lambda_{c}=\tilde{h}$ (see Ref.~\cite{GNF_2021} for the details).} {This technique allows us to overcome the divergence of the self-energies and reach the limit $h\to 0$ (see the inset in the lower panel of Fig.~\ref{SzHp1}). We find that in this approach for small $h$, the magnetization also exhibits a strictly linear magnetic field dependence and converges to a nonzero value in the limit $h\to 0$, which confirms the SDW ground state.} It should be noted that the counterterm technique, while giving the correct physical state, slightly overestimates the magnetization compared to the one obtained from the linear fit of the corresponding dependences shown in the main part of Fig.~\ref{SzHp1}. For this reason, we do not use this technique for quantitative analysis in our study.

Although for any fixed $N_{\rm vac} $ one can find configurations that belong to either the SM or SDW phase,
the quantitative distribution between these two phases is strongly dependent on $N_{\rm vac}$. To show this, we considered $S_{\rm st}^{(N_{\rm vac})}(h)$ dependencies for $150$ random configurations for each $N_{\rm vac}=2,4,6,8$, and attributed each configuration either to the SM state
or to the SDW state.
The resulting distribution between the SM and SDW phases is shown in Fig.~\ref{SM_SDW_distr}. For $N_{\rm vac}=2$, the SM state dominates over the SDW state, which means that it is more likely to find a random system with this number of vacancies in the SM state. However, already for $N_{\rm vac}=4$, the situation changes and the SDW state becomes more preferable (in a probabilistic sense). With increasing of $N_{\rm vac}$, this difference becomes even more pronounced and for $N_{\rm vac}=8$ the realization of the SDW state has the highly superior probability.

\section{Characteristic effects of disorder}
To estimate the overall effect of the vacancies, we generate $n$ different random configurations of the system with $N_{\rm vac}$ vacancies and calculate average values $\langle X^{(N_{\rm vac})}\rangle_{n}$ over these configurations (samples), where $X\in\{S_{\rm st}, \Delta_{\rm st}, G\}$. 
To consider the limit of $h \to 0$, $\langle X^{(N_{\rm vac})}\rangle_{n}$ are calculated for five different values of the magnetic field from $h=h_{m}$ to $h=0.2 h_{m}$ with the step $0.2 h_{m}$, where $h_{m}=0.0185t$ is the highest magnetic field used in the present study.
{The typical} dependence of the disorder-averaged relative staggered magnetization $\langle S_{\rm st}^{(N_{\rm vac})} \rangle_{n}$ on the number of samples $n$ ($n\leq 150$) is plotted in Fig.~\ref{S_vac_n_p1} {for $h{\leq}h_{m}$} and $N_{\rm vac}\leq 8$. The averaging over a small number of samples ($n\lesssim 50$) produces relatively large fluctuations of $\langle S_{\rm st}^{(N_{\rm vac})}\rangle_{{n}}$. These fluctuations are caused by the difference in the relative staggered magnetizations of different configurations. However, one can see that for sufficiently large $n$ ($n\gtrsim 100$), $\langle S_{\rm st}^{(N_{\rm vac})}\rangle_{n}$ is almost independent of $n$. 
Similarly, we find that {for $n\gtrsim 100$} the disorder-averaged linear conductance $\langle G^{(N_{\rm vac})}\rangle_{n}$ also exhibits only weak fluctuations and the disorder-averaged relative difference between the occupation number of the sublattices $\langle \Delta_{\rm st}^{(N_{\rm vac})}\rangle_{n}$ is almost independent of $n$. It is important to note that for a subset of considered configurations, due to strong SDW correlations, the electron-electron interaction vertices {(as well as the self-energy)} exhibit divergences in the region of small magnetic fields, $0<h\lesssim 0.4h_{m}$. This tendency becomes more pronounced as $N_{\rm vac}$ increases. For these configurations, we extract $X^{(N_{\rm vac})}$ values for small magnetic fields by extrapolating fRG data obtained for higher ones. \par
Figure~\ref{S_vac_n_p1} also shows $\langle S_{\rm st}^{(N_{\rm vac}=2)}\rangle_{n}$ at sufficiently large $n$ for various magnetic fields. The obtained dependencies are qualitatively similar to those obtained at $h=h_m$, up to the decrease of magnetization, related to the decrease of magnetic field.
It is important to note that the disorder-averaged relative staggered magnetization shows stable convergence to almost constant values for $n\gtrsim 100$ for all considered magnetic fields. For the other vacancy concentrations considered, the results are qualitatively similar to the ones presented in {Fig.~\ref{S_vac_n_p1}}. In general, for a fixed $N_{\rm vac}$, we find that $n\gtrsim 100$ configurations are sufficient to achieve a regime in which the disorder-averaged values have only small deviations from each other.\par

\begin{figure}[h!]
\centering
\includegraphics[width=0.98\linewidth]{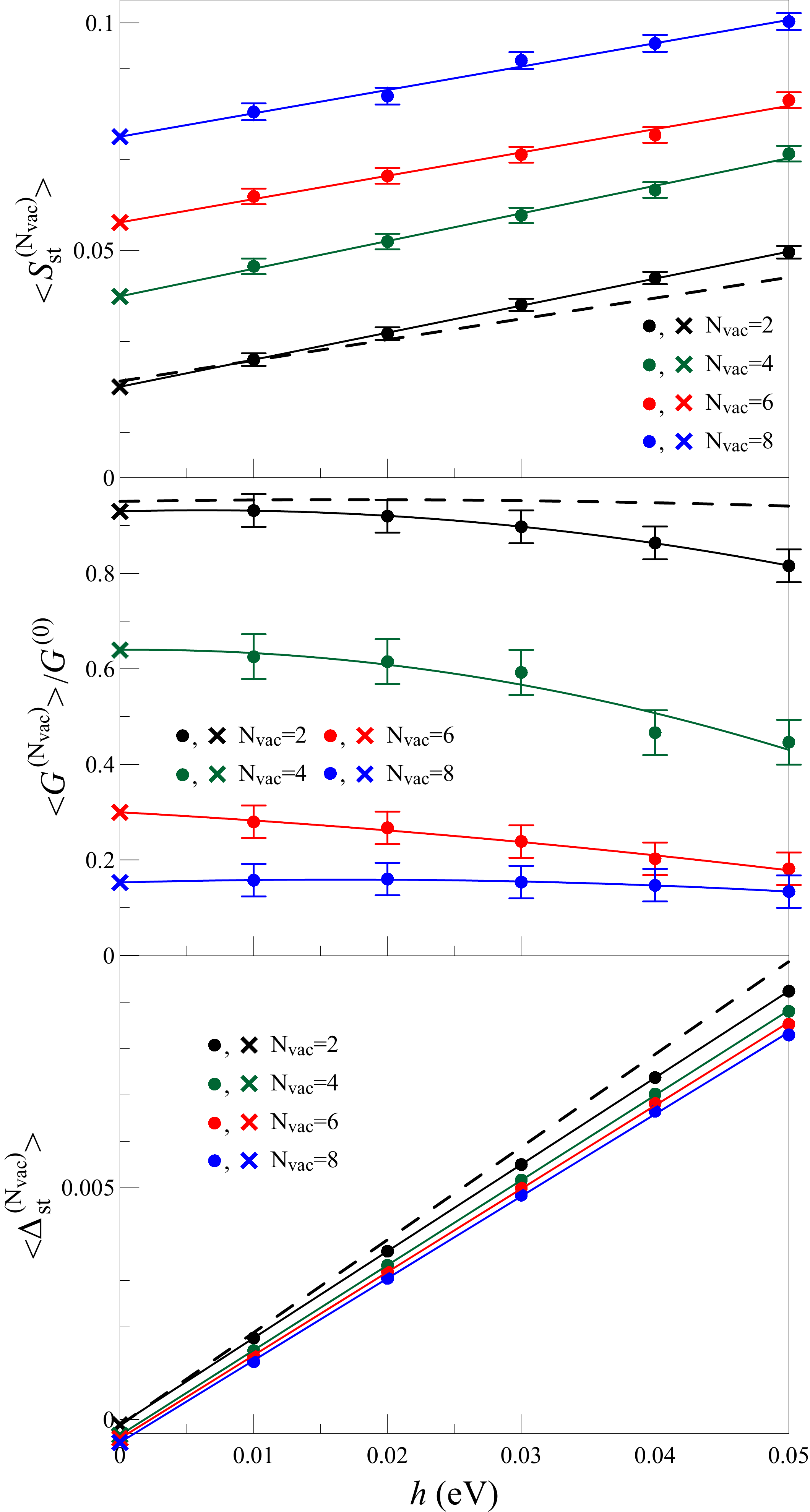}
\caption{(Color online) {The characteristic disorder-averaged relative staggered magnetization $\langle S_{\rm st}^{(N_{\rm vac})}\rangle$ (upper panel), linear conductance $\langle G^{(N_{\rm vac})}\rangle/G^{(0)}$ (middle panel), and the relative difference between the occupation of the sublattices $\langle \Delta_{\rm st}^{(N_{\rm vac})}\rangle$ (lower panel) as a function of magnetic field $h$ for various $N_{\rm vac}$.} {The lines in the upper and lower (or middle) panel are linear (or quadratic)} fits to the fRG data (circles), the cross symbols represent values obtained by extrapolation to $h=0$. $G^{(0)}$ is the value of the linear conductance for $N_{\rm vac}=h=0$. The dashed lines show the results corresponding to the hexagonal GNF system with armchair edges ($N_{\rm at}=114$). The horizontal bars show the standard errors of the mean (SEM) of the disorder-averaging procedure.
} 
\label{G_h_average}
\end{figure} 
We estimate the characteristic disorder-averaged values 
in the presence of $N_{\rm vac}$ vacancies in the system by 
$\langle X^{(N_{\rm vac})}\rangle_{}=\langle X^{(N_{\rm vac})}\rangle_{150}$. 
{
Results for $\langle X^{(N_{\rm vac})}(h)\rangle_{}$ for different numbers of vacancies $N_{\rm vac}$ are presented in Fig.~\ref{G_h_average}. One can see that linear extrapolation of the characteristic disorder-averaged relative staggered magnetization $\langle S_{\rm st}^{(N_{\rm vac})}\rangle$ to $h=0$ gives nonzero values for all considered $N_{\rm vac}$. In contrast, we do not find appreciable charge density wave order in the limit $h\rightarrow 0$ (i.e., $\langle \Delta_{\rm st}^{(N_{\rm vac})}\rangle\approx 0$ for $h \to 0$; see the lower panel of Fig.~\ref{G_h_average}), despite the symmetry breaking field $\delta=h$ is included in the calculation. The characteristic disorder-averaged linear conductance $\langle G^{(N_{\rm vac})}\rangle$ shown in Fig.~\ref{G_h_average} is nonlinear and displays the quadratic dependence on $h$. We estimate the values of $\langle G^{(N_{\rm vac})}\rangle$ for $h=0$ by quadratic extrapolation of the data for $h>0$.}\par

To reveal effects related to the geometry of the GNF edges, we also consider a GNF system with armchair edges consisting of $N_{\rm at}=114$ atoms in absence of vacancies. The position of the corresponding system with respect to the leads is identical to the one shown in Fig.~\ref{ConfVac0}. The results in this case for $\langle X^{(N_{\rm vac})}(h)\rangle_{}$ for $N_{\rm vac}=2$ are shown in Fig~\ref{G_h_average} (see the dashed lines). One can see that both  $\langle S_{\rm st}^{(N_{\rm vac})}(h)\rangle$ and $\langle \Delta_{\rm st}^{(N_{\rm vac})}(h)\rangle$ dependencies are relatively close to the ones obtained for the zigzag-edge geometry case. As expected, the linear conductance is found to be more sensitive to changes in edge geometry and as a consequence the characteristic disorder-averaged linear conductance $\langle G^{(N_{\rm vac})}\rangle$ significantly deviates from the corresponding data for the GNF system with zigzag edges. However, we note that the presence of vacancies in this case also leads to clear suppression of the linear conductance. In general, one can see that the geometry of the edges does not lead to a qualitative change of the vacancy effects (at least for small vacancy concentrations).

To also analyze the effects of the spatial inhomogeneity of the magnetization due to the presence of the edges, we consider the relative staggered magnetizations associated with the sites located at the edges $S^{(N_{\rm vac})}_{e, \rm st}$ and in the center $S^{(N_{\rm vac})}_{c, \rm st}$ of the GNF system
\begin{equation}
S^{(N_{\rm vac})}_{e(c), \rm st} =N_{e(c)}^{-1}\sum_{j\in e(c)}|{\langle}n_{j,\uparrow}{\rangle}-{\langle}n_{j,\downarrow}{\rangle}|, 
\label{S_ec_2}
\end{equation}
\begin{figure}[t]
\centering
\includegraphics[width=1.0\linewidth]
{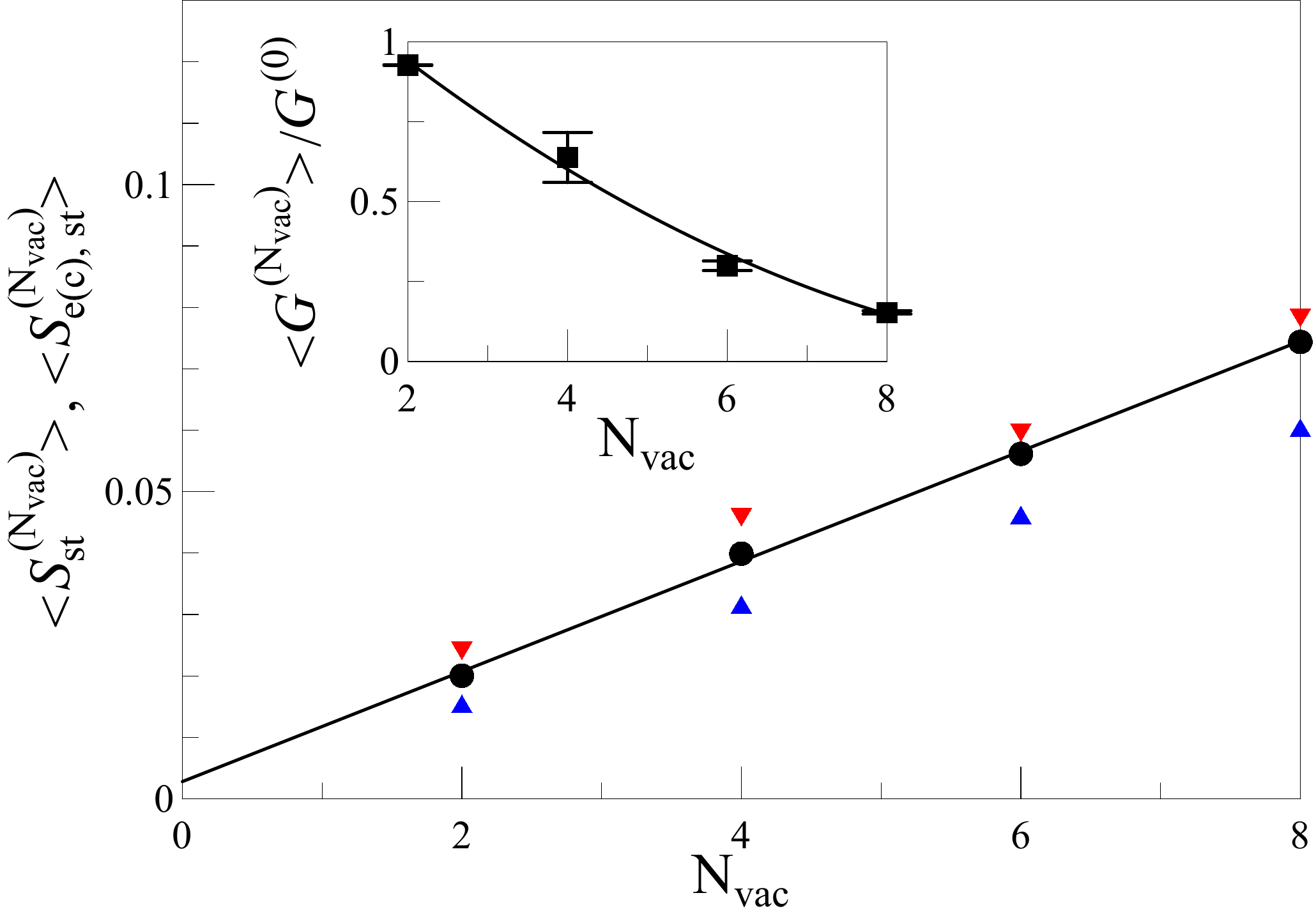}
\caption{{(Color online) The disorder-averaged staggered magnetization in the whole nanoflake $\langle S_{\rm st}^{(N_{\rm vac})}\rangle$ (circles) and that at the central (edge) sites $\langle S_{c(e), \rm st}^{(N_{\rm vac})}\rangle$ [upward (downward) triangles] for $h \to 0$ as a function of $N_{\rm vac}$. Inset: The disorder-averaged linear conductance $\langle G^{(N_{\rm vac})}\rangle/G^{(0)}$ for $h \to 0$ as a function of $N_{\rm vac}$ (the error bars represent the standard error of the fit of the data shown in the {middle} panel of Fig.~\ref{G_h_average}). The line in the main plot (inset) is a linear (quadratic) fit to the fRG data and $G^{(0)}\approx 6\times 10^{-4}G_{0}$ is the value of the linear conductance for $N_{\rm vac}=h=0$.}} 
\label{ch_val_150}
\end{figure}
where $N_{e(s)}$ is the number of edge (central) atoms in the GNF system (in the absence of vacancies $N_{e}=42$ and $N_{c}=6$) and the summation in Eq.~(\ref{S_ec_2}) is restricted to the edge (central) sites of the GNF. The corresponding relative staggered magnetization averaged over $n$ random configurations are given by $\langle S^{(N_{\rm vac})}_{e(c),\rm st} \rangle_{n}=({1}/{n})\sum_{k=1}^{n}{S^{(N_{\rm vac})}_{e(c), \rm st}({k})}$, where $S^{(N_{\rm vac})}_{e(c), \rm st}({k})$ is the value of $S^{(N_{\rm vac})}_{e(c), \rm st}$ for $k$th random disorder realization. We find that the disorder-averaged staggered magnetization corresponding to the edge sites is much stronger than that for the central sites. For all considered vacancy concentrations, the characteristic disorder-averaged staggered magnetization $\langle S_{\rm st}^{(N_{\rm vac})}\rangle$, which takes into account the contributions from all sites of the system, lies between the characteristic magnetizations $\langle S_{e,\rm st}^{(N_{\rm vac})}\rangle$ and $\langle S_{c, \rm st}^{(N_{\rm vac})}\rangle${, $\langle S_{e(c),\rm st}^{(N_{\rm vac})}\rangle=\langle S_{e(c),\rm st}^{(N_{\rm vac})}\rangle_{150}$. 
} 

The corresponding results for the characteristic disorder-averaged relative staggered magnetization $\langle S_{\rm st}^{(N_{\rm vac})}\rangle_{}$ and $\langle S_{e(c),\rm st}^{(N_{\rm vac})}\rangle$ {in the limit $h \to 0$} are presented in Fig.~\ref{ch_val_150}. It can be seen that $\langle S_{\rm st}^{(N_{\rm vac})}\rangle_{}$ increases approximately linearly with $N_{\rm vac}$.
Both the disorder-averaged sublattice magnetization in the center and at the edges increase approximately proportional to each other with an increase in the concentration of vacancies $N_{\rm vac}$, with the disorder-averaged ratio $\langle R^{(N_{\rm vac})}\rangle_{}=\langle S^{(N_{\rm vac})}_{e, \rm st}/S^{(N_{\rm vac})}_{c, \rm st}\rangle_{150}\simeq 1.5$, which weakly depends on the concentration of vacancies.


In Fig.~\ref{ch_val_150}, we also show the dependence of the characteristic disorder-averaged linear conductance $\langle G^{(N_{\rm vac})}\rangle_{}$ on the number of vacancies $N_{\rm vac}$. One can see that the conductance shows only a weak nonlinear (quadratic) dependence on $N_{\rm vac}$ and thus, approximately linearly decreases with increasing number of vacancies. For $N_{\rm vac}{=8}$, the conductance is suppressed by disorder significantly (to $\langle G^{(N_{\rm vac})}\rangle_{}\approx {0.16} G^{(0)}$) compared to the conductance $G^{(0)}$ for the system without vacancies. The decrease of the linear conductance correlates with the increase of the relative staggered magnetization (a similar correlation was previously noted for GNFs without vacancies~\cite{GNF_2021}). 

\section{Conclusions} 
The presence of vacancies in the GNF with screened realistic long-range electron interactions strongly modifies its magnetic and transport properties: (i) while GNFs with no vacancies are nonmagnetic for the realistic Coulomb interactions, the presence of vacancies yields to a strong enhancement of the magnetic correlations, which can be strong enough for GNF to be in the SDW state; (ii) the probability of GNF with a random configuration of vacancies to be in the SDW (SM) ground state gradually increases (decreases) with increase of vacancy concentration; (iii) the linear conductance shows a weak nonlinear dependence on the vacancy concentration and decreases with its increase.

The functional renormalization group approach used in the present study allows for a clear recognition and characterization of the magnetic/charge phases of GNF systems with vacancies. 
These capabilities are combined together with the ability to relatively quickly scan through large sets of arbitrary configurations with realistic long-range Coulomb interaction. In this sense, the extension of this method to more complicated models of disorder and other graphene nanosystems, such as graphene nanotubes with disorder, seems quite feasible and is of considerable interest.

\section*{Acknowledgments} 
The work was performed within the state assignment from the Ministry of Science and Higher Education of Russia (Theme “Quant” 122021000038-7) and partly supported by RFBR Grant No. 20-02-00252a. A.A.K. also acknowledges the financial support from the Ministry of Science and Higher Education of the Russian Federation (Agreement No. 075-15-2021-606). The calculations were performed on the Uran supercomputer at the IMM UB RAS.
\end{document}